\documentclass[10pt,A4paper,conference]{IEEEtran}
\usepackage{amsmath}
\usepackage{amssymb}
\usepackage{upref}
\usepackage{amsfonts}
\usepackage{graphicx}
\usepackage{epic}
\usepackage{eepic}
\usepackage{psfrag}
\usepackage{verbatim}

\newcommand{\deff}{\mbox{$\stackrel{\rm def}{=}$}}

\newcommand{\field}[1]{\mathbb{#1}}

\newcommand{\Z}{\field{Z}}

\newcommand{\cE}{{\cal E}}

\newcommand{\cA}{{\cal A}}
\newcommand{\cB}{{\cal B}}
\newcommand{\cC}{{\cal C}}
\newcommand{\cG}{{\cal G}}
\newcommand{\cL}{{\cal L}}

\newcommand{\cS}{{\cal S}}

\newcommand{\cP}{{\cal P}}

\newcommand{\sP}{\cP}
\newcommand{\sG}{\cG}

\newcommand{\Gr}{\smash{{\sG\kern-1.5pt}_q\kern-0.5pt(n,k)}}
\newcommand{\Gfourk}{\smash{{\sG\kern-1.5pt}_q\kern-0.5pt(4k,2k)}}
\newcommand{\Gk}{\smash{{\sG\kern-1.5pt}_q\kern-0.5pt(n,k_1)}}
\newcommand{\Gkk}{\smash{{\sG\kern-1.5pt}_q\kern-0.5pt(n,k_2)}}
\newcommand{\Grtwo}{\smash{{\sG\kern-1.5pt}_2\kern-0.5pt(n,k)}}
\newcommand{\Gkone}{\smash{{\sG\kern-1.5pt}_q\kern-0.5pt(n,k_1)}}
\newcommand{\Gktwo}{\smash{{\sG\kern-1.5pt}_q\kern-0.5pt(n,k_2)}}
\newcommand{\Ps}{\smash{{\sP\kern-2.0pt}_q\kern-0.5pt(n)}}

\newcommand{\bv}{{\bf v}}

\newtheorem{theorem}{Theorem}
\newtheorem{cor}{Corollary}
\newtheorem{lemma}[theorem]{Lemma}

\begin{document}

\title{On the Existence of Perfect Codes\\ for Asymmetric Limited-Magnitude Errors}

\author{\authorblockN{Sarit Buzaglo}
\authorblockA{Dept. of Computer Science\\
Technion-Israel Institute of Technology\\
Haifa 32000, Israel \\
Email: sarahb@cs.technion.ac.il} \and
\authorblockN{Tuvi Etzion}
\authorblockA{Dept. of Computer Science\\
Technion-Israel Institute of Technology\\
Haifa 32000, Israel \\
Email: etzion@cs.technion.ac.il}}

\maketitle
\begin{abstract}
Block codes, which correct asymmetric errors with
limited-magnitude, are studied. These codes have been applied
recently for error correction in flash memories. The codes will be
represented by lattices and the constructions will be based on a
generalization of Sidon sequences. In particular we will consider
perfect codes for these type of errors.
\end{abstract}

\section{Introduction}
\label{sec:introduction}

Asymmetric error-correcting codes were subject to extensive
research due to their application in coding for computer
memories~\cite{RaFu89}. The advance of technology and the
appearance of new nonvolatile memories, such as flash memory, led
to a new type of asymmetric errors which have limited-magnitude. A
multilevel flash cell is electronically programmed into $q$
threshold levels which can be viewed as elements of the set $\{
0,1, \ldots , q-1 \}$. Errors in this model are usually in one
direction and are not likely to exceed a certain limit. This means
that a cell in level $i$ can be raised by an error to level~$j$,
such that $i < j$ and $j-i \leq \ell$, where $\ell$ is the error
limited-magnitude.

Asymmetric error-correcting codes with limited-magnitude were
proposed in~\cite{AAK02} and were first considered for nonvolatile
memories in~\cite{CSBB07,CSBB}. Recently, several other papers
have considered the problem, e.g.~\cite{Dol10,ElBo10,KBE11,YSVW}.

In this work we mainly consider perfect codes for asymmetric
limited-magnitude errors. We will consider only linear codes,
unless otherwise is stated. Each $t$-error-correcting perfect code
in the Hamming scheme, over GF($q$), is also a perfect code for
error-correction of $t$ asymmetric errors with limited-magnitude
$q-1$~\cite{CSBB}. Especially, a Hamming code of length
$n=\frac{q^r-1}{q-1}$, over GF($q$), can correct one asymmetric
error with limited-magnitude $q-1$. Additional perfect codes for
correction of one asymmetric error with limited-magnitude~$\ell$
are obtained from tiling of $\Z^n$ with semi-crosses whose arms
have length $\ell$~\cite{Ste86}. Perfect unbalanced
limited-magnitude codes were considered in~\cite{Sch11}.

The rest of this work is organized as follows. In
Section~\ref{sec:basic} we will define the basic concepts for
codes which correct $t$ asymmetric errors with limited-magnitude
$\ell$. We will show a convenient way to handle such codes and
discuss three equivalent representations of such codes. In
Section~\ref{sec:construct} we will present a new construction for
perfect codes of length $n$ which correct $n-1$ asymmetric errors
with limited-magnitude~$\ell$, for any given $\ell$. In
Section~\ref{sec:nonexist} we show that perfect codes of length
$n$ which correct $n-2$ asymmetric errors with limited-magnitude
one cannot exist. We conclude in Section~\ref{sec:conclude}.

\section{Basic Concepts}
\label{sec:basic}

For a word $X=(x_1,x_2,\ldots , x_n) \in Q^n$, the \emph{Hamming
weight} of $X$, $w_H(X)$, is the number of nonzero entries in $X$,
i.e., $w_H(X) = \left| \{ i~:~ x_i \neq 0 \} \right|$.

A \emph{code} $\cC$ of \emph{length} $n$ over the alphabet $Q = \{
0,1, \ldots , q-1 \}$ is a subset of $Q^n$. A vector $\cE
=(e_1,e_2,\ldots,e_n)$ is a \emph{$t$-asymmetric-error with
limited-magnitude $\ell$} if $w_H (\cE) \leq t$ and $0 \leq e_i
\leq \ell$ for each $1 \leq i \leq n$. The sphere $\cS(n,t,\ell)$
is the set of all $t$-asymmetric-errors with limited-magnitude
$\ell$. A code $\cC \subseteq Q^n$ can correct $t$ asymmetric
errors with limited-magnitude $\ell$ if for any two codewords
$X_1,~X_2$, and any two $t$-asymmetric-errors with
limited-magnitude $\ell$, $\cE_1,~\cE_2$, such that $X_1 + \cE_1
\in Q^n$, we have that $X_1 + \cE_1 \neq X_2 + \cE_2$.

For simplicity it is more convenient to consider the code~$\cC$ as
a subset of $\Z_q^n$, where all the additions are performed
modulo~$q$. Such a code $\cC$ can be viewed also as a subset of
$\Z^n$ formed by the set $\{ X + qY ~:~ X \in \cC,~ Y \in \Z^n
\}$.

We will represent a linear code $\cC$, over $\Z_q^n$, which
corrects $t$ asymmetric errors with limited-magnitude $\ell$, in
two more different ways. The first is by an integer lattice and
the second is by a generalization of the well-known Sidon
sequence, the $\cB_h$ sequence. We will show an equivalence
between the three representations.

An \emph{integer lattice} $\Lambda$ is an additive subgroup of
$\Z^n$. We will assume that
\begin{equation*}
\Lambda \deff \{ u_1 \bv_1 + u_2 \bv_2 + \cdots + u_n \bv_n ~:~
u_1,u_2, \cdots,u_n \in \Z \}
\end{equation*}
where $\{ \bv_1, \bv_2,\ldots, \bv_n \}$ is a set of linearly
independent vectors in $\Z^n$. The set of vectors $\{ \bv_1,
\bv_2,\ldots, \bv_n\}$ is called \emph{the basis} for $\Lambda$,
and the $n \times n$ matrix
$$
{\bf G} \deff \left[\begin{array}{cccc}
v_{11} & v_{12} & \ldots & v_{1n} \\
v_{21} & v_{22} & \ldots & v_{2n} \\
\vdots & \vdots & \ddots & \vdots\\
v_{n1} & v_{n2} & \ldots & v_{nn} \end{array}\right]
$$
having these vectors as its rows is said to be the \emph{generator
matrix} for $\Lambda$.

The {\it volume} of a lattice $\Lambda$, denoted by $V( \Lambda
)$, is inversely proportional to the number of lattice points per
a unit volume. There is a simple expression for the volume of
$\Lambda$, namely, $V(\Lambda)=| \det {\bf G} |$.

A set $P \subseteq \Z^n$ is a \emph{packing} of $\Z^n$ with a
shape $\cS$ if copies of $\cS$ placed on the points of $P$ (in the
same relative position) are disjoint. A set $T$ is a \emph{tiling}
of $\Z^n$ if it is a packing and the disjoint copies of $\cS$
cover $\Z^n$. A lattice $\Lambda$ is a \emph{lattice packing
(tiling)} with the shape $\cS$ if $\Lambda$ forms a packing
(tiling) with $\cS$. The following lemma is well known.

\begin{lemma}
A necessary condition that the lattice $\Lambda$ defines a lattice
packing (tiling) with the shape $\cS$ is that $V(\Lambda) \geq
|\cS|$ ($V(\Lambda)=|\cS|$), where $|\cS|$ denote the volume of
$\cS$.
\end{lemma}

A linear code $\cC$, over $\Z_q^n$, which corrects $t$ asymmetric
errors with limited-magnitude $\ell$ viewed as a subset of $\Z^n$
is equivalent to an integer lattice packing with the shape
$\cS(n,t,\ell)$. Therefore, we will call this lattice a
\emph{lattice code}.

Let $\cA(n,t,\ell)$ denote the set of lattice codes in $\Z^n$
which correct $t$ asymmetric errors with limited-magnitude $\ell$.
A code $\cL \in \cA(n,t,\ell)$ is called \emph{perfect} if it
forms a lattice tiling with the shape $\cS(n,t,\ell)$.

Let $[\ell]$ be the set $\{0,1,2,...,\ell\}$ and let $G$ be an
Abelian group. A $\cB_h[\ell](G)$ sequence of length $m$ is a
sequence (set) of $m$ elements in $G$, $b_1,b_2,...,b_m$ ($\{
b_1,b_2,...,b_m \}$) such that all sums
$$
\sum_{j=1}^{m}\alpha_j b_j~,
$$
where $\alpha_j\in [\ell]$ and at most $h$ of the the $\alpha_j$'s
are nonzero, are distinct elements of $G$. $\cB_h$ sequences were
first mentioned in~\cite{KBE11} for correction of asymmetric
errors with limited-magnitude.

\begin{lemma}
\label{lemm:codeToBh} If $\cL \in \cA(n,t,\ell)$ then there exists
an Abelian group $G$ of order $|G|=V(\cL)$ and a $\cB_t[\ell](G)$
sequence of length $n$.
\end{lemma}
\begin{lemma}
\label{lemm:BhToCode} Let $G$ be an Abelian group and let
$b_1,...,b_n$ be a $\cB_t[\ell](G)$ sequence. Then there exists a
lattice code $\cL \in \cA(n,t,\ell)$ with $V(\cL)\leq |G|$.
\end{lemma}
\begin{cor}
\label{cor:per_lattice} A perfect lattice code $\cL\in
\cA(n,t,\ell)$ exists if and only if there exists an abelian group
$G$ of order $|\cS(n,t,\ell)|$ and a $\cB_k[\ell](G)$ sequence of
length $n$.
\end{cor}

To form a code $\cC \subseteq \Sigma^n$, where $\Sigma \deff \{
0,1,\ldots , \sigma -1 \}$, which corrects $t$ asymmetric errors
with limited-magnitude $\ell$, one can take a lattice code $\cL\in
\cA(n,t,\ell)$. Then $\cC \deff \left( X+ \cL \right) \cap
\Sigma^n$, where $X$ is any element of $\Z^n$ added to all the
elements of the lattice $\cL$, is an appropriate code. Note that
the code $\cC$ is usually not linear.

\section{Perfect Codes which Correct $n-1$ Errors}
\label{sec:construct}

To use Corollary~\ref{cor:per_lattice}, we have to compute
$\cS(n,t,\ell)$.

\begin{lemma}
$\left| \cS(n,t,\ell) \right| = \sum_{i=0}^t \binom{n}{i} \ell^i$.
\end{lemma}
\begin{cor}
$\left| \cS(n,n-1,\ell ) \right| = (\ell +1)^n - \ell^n$.
\end{cor}

For the ring $G=\Z_q$, the ring of integers modulo $q$, let $G^*$
be the multiplicative group of $G$ formed from all the elements of
$G$ which have multiplicative inverses in $G$.

\begin{lemma}
\label{lemma:l_property} Let $n \geq 2$, $\ell \geq 1$, be two
integers and let $G$ be the ring of integers modulo
$(\ell+1)^n-\ell^n$, $\Z_{(\ell+1)^n-\ell^n}$. Then,

\begin{itemize}
\item[(P1)] $\ell$ is an element of $G^*$.

\item[(P2)] The element $x=(\ell+1)\cdot \ell^{-1}$ of $G$ is an
element of $G^*$ of order $n$.

\item[(P3)] The expression $1+x+x^2+...+x^{n-1}$ equals to
\emph{zero} in~$G$.
\end{itemize}
\end{lemma}

(P1) and (P2) are important in the construction obtained from the
following theorem, while (P3) is important for its proof.

\begin{theorem}
\label{thm:mainT} For each $n \geq 2$ and $\ell \geq 1$, let
$x=(\ell+1)\cdot \ell^{-1}\in \Z^{^*}_{(\ell+1)^n-\ell^n}$. Then
the set $\{1,x,x^2,...,x^{n-1}\}$ is a
$\cB_{n-1}[\ell](\Z_{(\ell+1)^n-\ell^n})$ sequence.
\end{theorem}
\begin{cor}
For each $n \geq 2$ and $\ell \geq 1$ there exists a perfect
lattice code $\cL \in \cA(n,n-1,\ell)$.
\end{cor}

\section{Nonexistence of some Perfect Codes}
\label{sec:nonexist}

Recall that there exists a perfect lattice code $\cL \in
\cA(n,t,\ell)$ for various parameters with $t=1$. Such codes also
exist for $t=n$ and all $\ell \geq 1$ and for the parameters of
the Golay codes and the binary repetition code of odd length. In
Section~\ref{sec:construct} we proved the existence of such codes
for $t=n-1$ and all $\ell \geq 1$. Next, we ask whether such codes
exist for $t=n-2$? Unfortunately, if $t=n-2$ and $\ell=1$ such
codes cannot exist. The proof is based on the following lemma.

\begin{lemma}
\label{lem:perfect_cond} If there exists a perfect lattice code in
$\cA(n,n-2,l)$ then $|\cS(n,n-2,\ell)|$ divides
$(\ell+1)^{n-2}\cdot(\ell+1+\alpha\cdot(n-2-\ell))$ for some
integer $\alpha$, $0\leq \alpha \leq \ell$.
\end{lemma}
\begin{theorem}
There are no perfect lattice codes in $\cA(n,n-2,1)$ for all
$n>3$.
\end{theorem}

\section{Conclusion}
\label{sec:conclude}

We discussed three different equivalent ways to consider linear
codes which correct $t$ asymmetric errors with limited-magnitude
$\ell$. One of these ways was to consider $\cB_h$ sequences. We
presented a construction of $\cB_h$ sequences which result in
perfect codes of length $n$ for correction of $n-1$ asymmetric
errors with limited-magnitude $\ell$ for any given $\ell$. A
related nonexistence result for $n-2$ errors and limited-magnitude
one was given. It is a major research problem to prove whether
more such perfect codes exist.

\section{Note Added}
\label{sec:conclude}

After the Arxiv submission we became aware of the work
in~\cite{KLNY} which contains Theorem~\ref{thm:mainT}.

\section*{Acknowledgment}
This work was supported in part by the Israel Science Foundation
(ISF), Jerusalem, Israel, under Grant No. 230/08.

%
%

\end{document}